\documentclass[runningheads,a4paper]{llncs}

\pdfoutput=1
\usepackage{amssymb}
\usepackage{graphicx}
\usepackage{url}

\urldef{\mailsa}\path|adrien.devresse@cern.ch, furano@cern.ch|  
\newcommand{\keywords}[1]{\par\addvspace\baselineskip
\noindent\keywordname\enspace\ignorespaces#1}

\begin{document}

\mainmatter  

\title{Efficient HTTP based I/O on very large datasets for high performance computing with the libdavix library}

\titlerunning{Efficient HTTP based I/O}

%
%
\author{Adrien Devresse \and Fabrizio Furano}
%

\institute{CERN, European Organization for Nuclear Research ,\\
Geneva, Switzerland \\
\mailsa\\
\url{http://cern.ch}}

\maketitle

\begin{abstract}
Remote data access for data analysis in high performance computing is commonly done with specialized data access protocols and storage systems. These protocols are highly optimized for high throughput on very large datasets, multi-streams, high availability, low latency and efficient parallel I/O. 
The purpose of this paper is to describe how we have adapted a generic protocol, the Hyper Text Transport Protocol (HTTP) to make it a competitive alternative for high performance I/O and data analysis applications in a global computing grid: the Worldwide LHC Computing Grid.
In this work, we first analyze the design differences between the HTTP protocol and the most common high performance I/O protocols, pointing out the main performance weaknesses of HTTP. Then, we describe in detail how we solved these issues. Our solutions have been implemented in a toolkit called davix, available through several recent Linux distributions. \\
Finally, we describe the results of our benchmarks where we compare the performance of davix against a HPC specific protocol for a data analysis use case.
\end{abstract}

\section{Introduction}

The Hyper Text Transport Protocol (HTTP)\cite{http_rfc_10} is today undoubtedly one of the most prevalent protocols on the internet. 

\par Initially created by Tim Berners-Lee for the World Wide Web at CERN in 1990\cite{www_tim}, HTTP is today much more than a simple protocol dedicated to HTML web page transport. The extensions of HTTP like WebDav\cite{webdav_rfc} or CalDav\cite{caldav_rfc}, the HTTP based protocols like UPnP or SOAP\cite{soap_publi} and the RESTful\cite{fielding_rest} architecture for Web Services have transformed HTTP into an universal versatile application layer protocol for the internet. The recent emergence of cloud computing\cite{cloud_book} and the popularization of big data storage\cite{big_data} based on RESTful Web services have definitively proved the universalism of HTTP. 

\par Today, HTTP is the foundation for interactions with commercial cloud storage providers like Amazon Simple Storage Service\cite{amazon_s3} or with Open Source Cloud Storage\cite{open_source_cloud} systems like OpenStack Swift \cite{openstack_cookbook} using REST API like S3\cite{amazon_s3_rest} or CDMI\cite{cdmi_rest}. HTTP is fully accepted as data transfer and data manipulation protocol in NoSQL databases and distributed storage systems in the Web World.
The association of REST APIs with the HTTP protocol usage offers a simple, standard, extensible, portable alternative to the legacy data access and file manipulation protocols or to the proprietary protocol of most distributed file systems. 

\par However, the popularity of HTTP was still not penetrating the High Performance Computing world. HPC\footnote{High Performance Computing} data access have highly specific and strict requirements: very large data manipulation, low-latency, high-throughput, high-availability, highly parallel I/O and high-scalability. For these reasons, those use cases traditionally rely on highly specific systems and protocols adapted to such constrains.
The IBM GPFS\cite{gpfs_paper} protocol, the Lustre parallel distributed file system protocol, the Hadoop HDFS\cite{hadoop_guide} streaming protocol, the gridFTP protocol\cite{gridFTP_2007} or the XRootD protocol\cite{big_data} are widely used in super computing and grid computing environments. 

\par The focus of this work is to be able to make the HPC world benefit of all the momentum coming from the HTTP Ecosystem, like the RESTful and Cloud Storage services, by creating a high performance I/O layer based on the HTTP protocol, compatible with standard services and competitive with the HPC I/O specific protocols. \\
To achieve this, we have created \textit{libdavix}\cite{davix_website}\cite{davix_project}, an I/O layer implementation optimized for data analysis and HPC I/O in distributed environments.

\section{Background and Related work}

\subsection{HTTP as a data management and data transfer protocol}

The stateless nature of HTTP, associated with the atomic nature of its primitives, provides a simple, reliable and powerful consistency model in distributed environment. The success and the major diffusion of the RESTful architecture introduced by Roy Fielding\cite{fielding_rest} for the Web World illustrate perfectly this \cite{http_rfc_11}.
 
\par The HTTP PUT method provides an object level idempotent write operation that can be used for an atomic resource creation or a resource content update. The basic HTTP GET method gives a safe, cacheable, atomic and idempotent Object level read operation and can be used to access and retrieve safely a remote resource. These two methods, associated with the HTTP DELETE method, are enough to satisfy the four basic functions Create, Retrieve, Update, Delete (CRUD)\cite{crud_james_martin} of any basic persistent storage system \cite{web2_rest_crud}. 

\par The properties of the HTTP protocol make it suitable for data transfer in a distributed environment and easily justify the emergence of persistency and data storage solutions using RESTful interfaces. It is the case for instance of the NoSQL database couchdb\cite{couchdb_guide}, of the NoSQL key-store Ryak\cite{riak_db}, of the distributed file system HDFS with httpFS\cite{hadoop_guide}, of the Amazon Simple Storage Service (Amazon S3)\cite{amazon_s3} or of any similar RESTful Object Storage service. \\
Again, the universalism of HTTP and the quality of its ecosystem associated with its simple and flexible design, makes it a first quality choice for a generic data transfer protocol today.

\subsection{Efficient parallel I/O operations}

Intensive data analysis applications requires high degree of I/O parallelism, robustness over large transfers and low I/O latency. A high energy physics application typically processes in parallel a very large number of events from different files located in large distributed data stores, triggering a significant number of parallel I/O operations. 

\par For such use cases in, the grid computing models in the HEP community use a mix of I/O frameworks for HPC\footnote{High performance computing} data access, the XRootD framework\cite{xrootd_paper}, the GridFTP protocol\cite{gridFTP_2007} with the Globus toolkit\cite{foster1999globus}, HDFS of Hadoop\cite{hadoop_guide} and IBM GPFS\cite{gpfs_paper}. \\
All these frameworks are highly optimized for parallel access, high throughput and efficient I/O scheduling of multiple requests. The XRootD framework implements its own I/O scheduler, it supports parallel asynchronous data access on top of its own I/O multiplexing.\\
The GridFTPv2  protocol has separated control and data channels and supports multiple data streams from different data sources on top of TCP or UDP.\\
The HDFS architecture is specially designed for large file storage, high throughput, hot file replications and data streams from multiple DataNodes. 

\par To explore the possibilities for a solution based on HTTP, we defined a set of criteria to meet.
\begin{itemize}
  \item Efficiency for large data transport and parallel I/O.
  \item Compatibility with existing network infrastructure and services.
  \item Low I/O latency: avoid useless handshakes, useless reconnections and redirections. 
\end{itemize}

\par The original design of HTTP did not match very well these points. \\
The HTTP 1.0 standard recommends the usage of one TCP connection per request to the server. This approach has been already proven inefficient due to the TCP slow start mechanism \cite{http_perf_problem}. Executing a HEP\footnote{High energy physics} data analysis work-flow with a very large number of parallel small sized requests in such conditions would lead to a major performance penalty. 
\par To mitigate the effect of this behaviour, HTTP 1.1 introduced the persistent connection support with the \textit{KeepAlive} mechanism and  the support of request pipelining over the same connection \cite{http_rfc_11}. 

\begin{figure}[h]
  \centering
  \includegraphics[width=8cm,height=5cm]{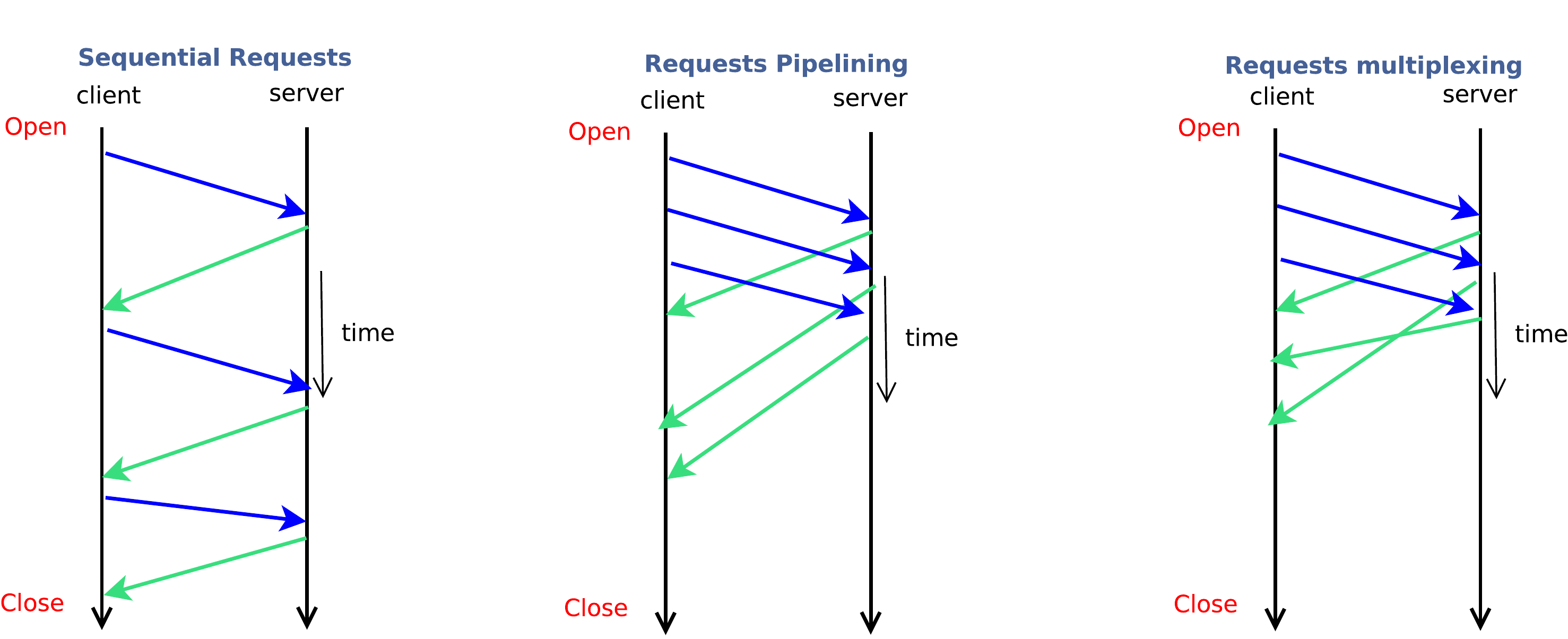}
  \caption{HTTP Request pipelining and Request Multiplexing}   
\end{figure}

\par However, HTTP pipelining suffers of several problems. 
Contrary to a protocol supporting modern multiplexing, the HTTP pipelining suffers of the HOL\footnote{Head of Line} problem.
The HTTP standard specifies that the treatment of a group of pipelined HTTP requests has to be processed in order. With such a requirement, any request pipelined suffering of a delay will cause a delay for all the following requests \cite{spdy_faq}. This is a unacceptable performance penalty in case of parallel I/O requests with different sizes. \\
The HTTP pipelining also suffers from other problems. It suffers of side effects with the TCP's nagle algorithm\cite{http_pipelining_nagle}. It often suffers of performance degradation due to aggressive pipeline interruptions with some servers implementations and due to the fact that the pipelining is not respected by most of the proxy servers.
 For these reasons, most of the current web browsers (Chrome, Firefox and Internet explorer) web browser disable or do not support  the HTTP pipelining mechanism. \\

\par To resolve these problems inherent to HTTP 1.1, several proposal have been made:
\begin{itemize}
  \item the \textbf{SPDY protocol} is "an application-layer protocol for transporting content over the web, designed specifically for minimal latency". SPDY acts as a session layer between HTTP and TCP. It supports multiplexing, prioritization and header compression\cite{spdy_protocol}. SPDY is currently the most mature implementation of multi-plexing for HTTP and supported by a majority of modern web browsers.
  \item \textbf{HTTP over SCTP} proposes to use the SCTP multi-homing and multi-plexing features with HTTP\cite{http_over_sctp}. SCTP is a acronym for Stream Control Transmission Protocol, it provides a message-based alternative to TCP.
  \item \textbf{WebMux and HTTP-NG} proposes the usage of the MUX protocol as session protocol to provide multi-plexing for HTTP\cite{webmux_overview}\cite{smux_spec}.
\end{itemize}

None of these approaches are considered acceptable for a High performance I/O usage. 

\par The SPDY protocol explicitly enforces the usage of SSL/TLS for protocol negotiation purpose. TLS introduces a negative performance impact for big data transfers \cite{tls_performance} and introduces a handshake latency that can not be mandatory in High performance computing. 

\par The HTTP over SCTP proposal implies naturally to replace the TCP protocol by SCTP. Like any other protocol aiming to replace the level 4 of the OSI model, the SCTP protocol triggers several concerns about the compatibility with the existing network architecture, about the NAT-traversals capability and about the support in old operating systems. 

\par The MUX, renamed WebMUX protocol, defined in 1998, focuses on an object oriented approach with in mind the support for technology like RMI\footnote{Remote Method Invocation}, DCOM\footnote{Distributed Component Object Model} or CORBA\footnote{Common Object Request Broker Architecture} which is not our use cases. Moreover, it has currently not been implemented in any major HTTP server.  

\par
\begin{figure}[h]
   \centering
   \includegraphics[width=7cm,height=7cm]{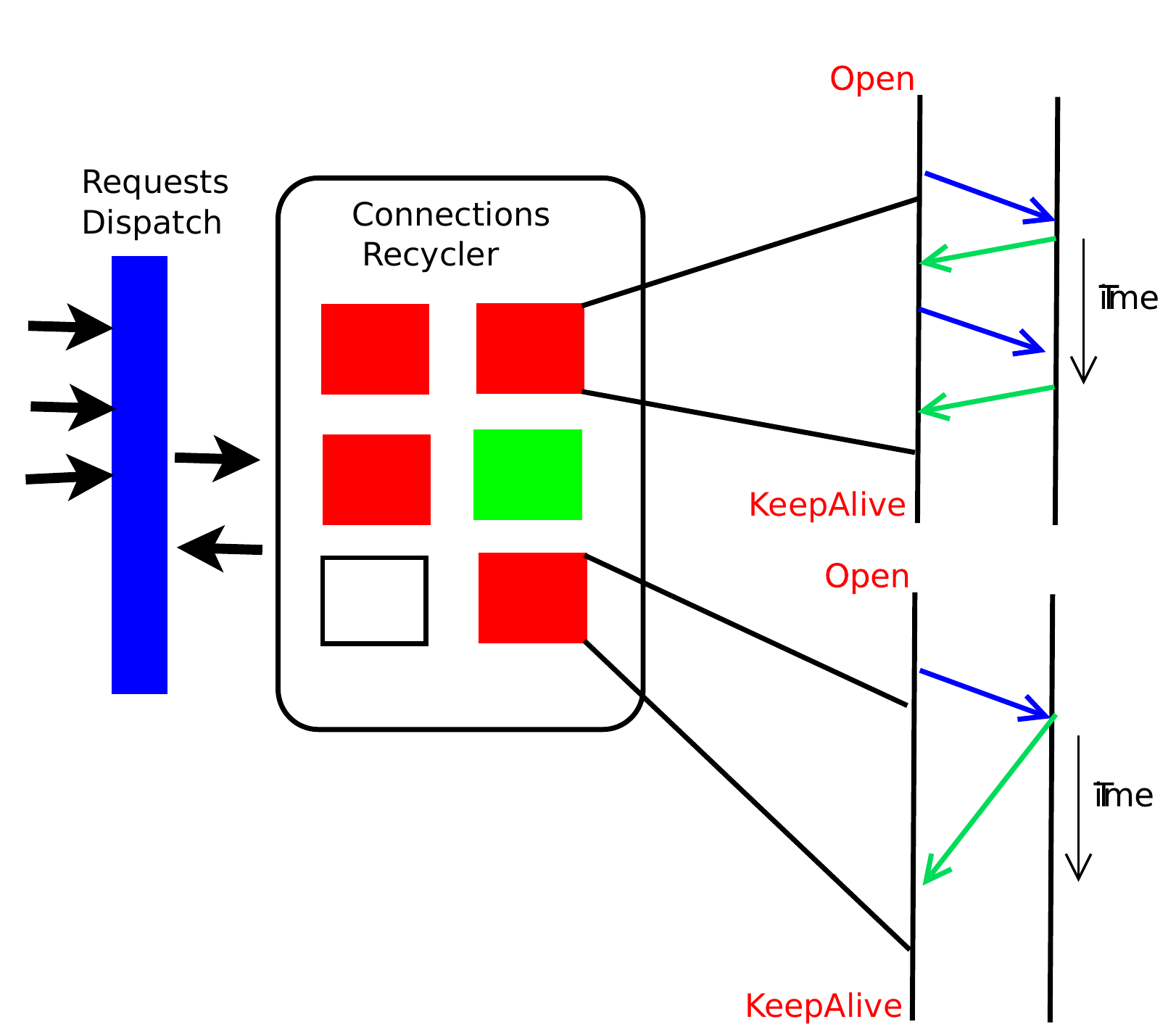}
   \label{fig:libdavix_parallelism}
   \caption{Dynamic connection pool with a thread-safe query dispatch}   
\end{figure}

 To satisfy our use cases, we adopted and implemented a different approach into \textit{libdavix} \cite{davix_website}\cite{davix_project}. We use a hybrid solution based on a dynamic connection pool with a thread-safe query dispatch system and a session recycling mechanism (See Figure-\ref{fig:libdavix_parallelism}). \\
Associated with the pool, we enforce an aggressive usage of the HTTP KeepAlive feature, \textit{libdavix} to maximize the re-utilization of the TCP connections and to minimize the effect of the TCP slow start. 

\par This method gives several benefits compared to previously quoted solutions. \\
First, it is fully compatibility with the standard HTTP 1.1 and with existing services and infrastructure. \\
Second, we supports efficient parallel request execution for repetitive I/O operations without suffering of the problems that are specific to the classical HTTP pipelining. nor necessitating a protocol modification to support multi-plexing. \\

\par This dispatch approach is particularly adapted to a HPC data-analysis work-flow: a repetitive concurrent access to a limited set of hosts exploits at the best the session recycling and maximizes the lifetime of the TCP connections. 
However, contrary to a pure multi-plexing solution that aims to the usage of one TCP connection per host, our approach uses a connection pool whose size is proportional to the level of concurrency. Consequently, an important degree of concurrency can result in a more important server load compared to a multi-plexed solution like spdy due to the number of connections allocated per client.
\par However, this is not a big issue for our HPC use case, the support of "vectored queries" of \textit{libdavix} explained in section \ref{sec:vectored_io} reducing significantly the number of simultaneous concurrent requests.

\subsection{Scalable random-access I/O for partial file access with HTTP}
\label{sec:vectored_io}

Very large data sets in distributed environments are generally partitioned into separated data subset objects distributed in several storage nodes.  \\
In High Energy Physics, a data set generally contains a important number of particle events decomposed in ROOT\cite{root_framework} \textit{TTrees} of events and stored in different compressed files. \\

\par This approach allows an easy data distribution and simplify data replication. It facilitates the partitioning the data in a distributed environment like in the the World Wide LHC Computing Grid (WLCG)\cite{wlcg_site}. At the same time, a data analysis with this data model is I/O intensive and generates a very large number of individual data accesses operation to the storage. In order to extract a set of specific events spread in different remote data sets, a HEP Application needs to read a large number of small segments of data in different remote objects. \\
To reduce the number of I/O requests, and, hence, the impact of latency with such patterns, high performance computing I/O protocols implement Data Sieving algorithms, two phase I/O algorithms\cite{data_sieving} or sliding window buffering algorithms.  \\
To the best of our knowledge, no nowadays HTTP I/O toolkit before davix implemented similar I/O optimization strategies.  \\

\par
\begin{figure}[h]
   \centering  
   \includegraphics[width=5cm,height=7cm]{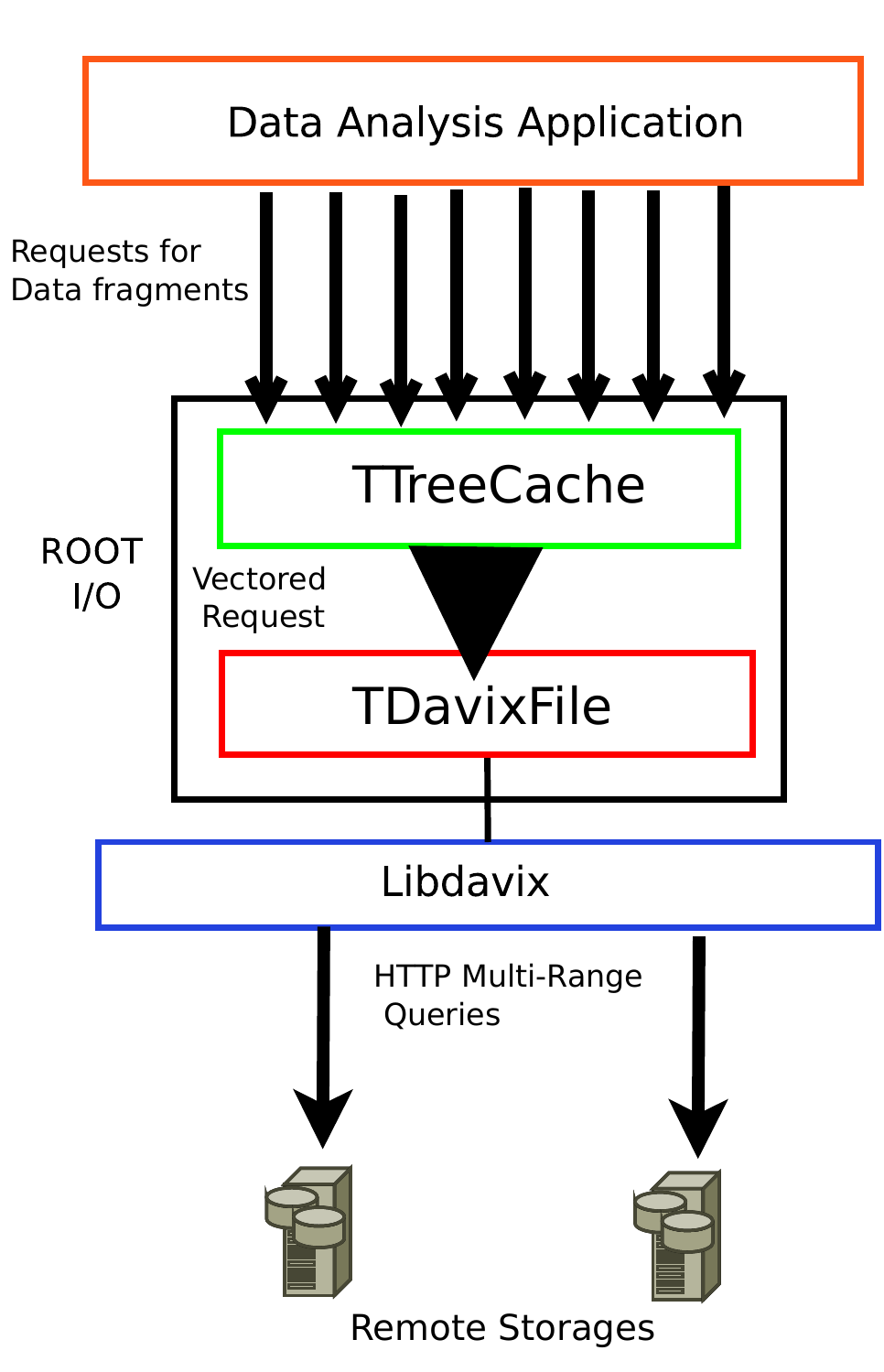}
   \label{fig:vectored_io}
   \caption{Vectored I/O requests support in the ROOT framework associated with LibDavix}   
\end{figure}

 We implemented in \textit{libdavix} a support for vectored packed operations with random position based on the multi-range feature of the HTTP 1.1 protocol (Figure \ref{fig:vectored_io}).\\
This feature allows to gather and pack a large number of fragmented random I/O requests directly in the ROOT\cite{root_framework} I/O framework via the TTreeCache\cite{ttreecache_atlas} in a large vectored query. Subsequently, this query is processed by \textit{libdavix} as one atomic remote I/O query.
This approach reduces drastically the number of remote network I/O operations and offers the advantage to reduce the necessity of parallel I/O operations and thus virtually eliminates the need for I/O multi-plexing.

\subsection{Multi-stream and multiple replicas I/O operations}

In grid computing computing, the unavailability of an input data required by a job is often the main cause of failure. Such situation leads to a redistribution of the data and a rescheduling of the job with a substantial performance impact on the execution time. \\
\par In an attempt to solve this issue, the XRootD protocol \cite{xrootd_paper} supports a federation mechanism to offer a more resilient access to a distributed resources. XRootD data servers can be federated hierarchically into a global virtual name-space. In case of unavailability of a resources in the closest data repository, the XRootD federation mechanism will locate a second available replica of this resource and redirect the client there.
\par HTTP alone does not support this feature. A classical HTTP access to an unavailable resource or a offline server will result in a I/O error. 
To improve the resiliency of the data layer, davix implements natively a support for the Metalink\cite{metalink_file} standard file format coupled to a fail-over and filtering mechanism for the offline replicas of a resource. 

\par A Metalink file is a standardized XML\cite{metalink_rfc} file containing several elements of meta-data information about an online resource: name, size, checksum, signature and location of the replicas of the resource. A Metalink file is a resource description and a set of ordered pointers to this resource. \\

Davix can use the Metalink information to apply two strategies: 
\begin{itemize}
  \item The \textbf{"fail-over" strategy} (default). In the case a resource is not available, davix try seamlessly to obtain the Metalink associated with this resource. Then \textit{libdavix} will try to access one per one the available replicas of this resource until being able to access the requested data on one of the available replica of the resource.
This approach improves drastically the resiliency of the data access layer and has the advantage to be without compromise or impact on the performances.

  \item The \textbf{"Multi-stream" strategy}. In this case, \textit{libdavix} will first try to obtain the Metalink of the resource and then proceed to a multi-source parallel download of each referenced chunk of data from a different replica. This approach has the advantage to maximize the network bandwidth usage on the client side and to offer the same resiliency improvement than the fail-over strategy. However, it has for main drawback to overload considerably the servers.
  
\end{itemize}

The combined usage of \textit{libdavix} for data analysis with a Replica catalogue or federation system supporting able to provide Metalink files like the DynaFed system (Dynamic Storage Federation)\cite{dynafed} enforces the global resilience of the I/O layer of any HPC application in a transparent manner. It provides the guarantee that a read operation on a resource will succeed as long as one replica of this resource is remotely accessible and referenced by the corresponding Metalink. 

\section{Performance analysis}

For our performance analysis, we executed a High energy analysis job based on ROOT framework\cite{root_framework} reading a fraction or the totality of around 12000 particles events from a 700 MBytes root file. This tests has been executed using both the XRootD framework and our libdavix solution as I/O layer. \\

\par Each test execution has been executed on WLCG through the Hammerloud Grid\cite{hammercloud} performance testing framework. \\
The execution of the test is always done on a standard Worder Node configuration of WLCG. \\
For both XRootD and davix, each test is run against the same server instance with the following configuration: Disk Pool Manager(DPM) Storage system, 4 Core Intel Xeon CPU, 12 GBytes of RAM, Scientific Linux CERN 6, 1 GB/s network link. \\
Two test executions are separated by 20 minutes. \\
Our statistics are based on an average of 576 tests executions over a period of 12 days. \\

We compare here the global execution time of the analysis job over different network configurations:

\begin{itemize}
  \item \textbf{LAN}: accessing the file over a gigabit Ethernet with low latency (latency \textless 5 ms)
  \item \textbf{PAN-European network}: accessing the file over GEANT\cite{geant} between Switzerland and UK (latency \textless 50 ms)
  \item \textbf{WAN}: accessing the file over Internet between Switzerland and USA (latency \textless 300 ms)
\end{itemize}

\begin{figure}[h]
   \includegraphics[width=16cm,height=8cm]{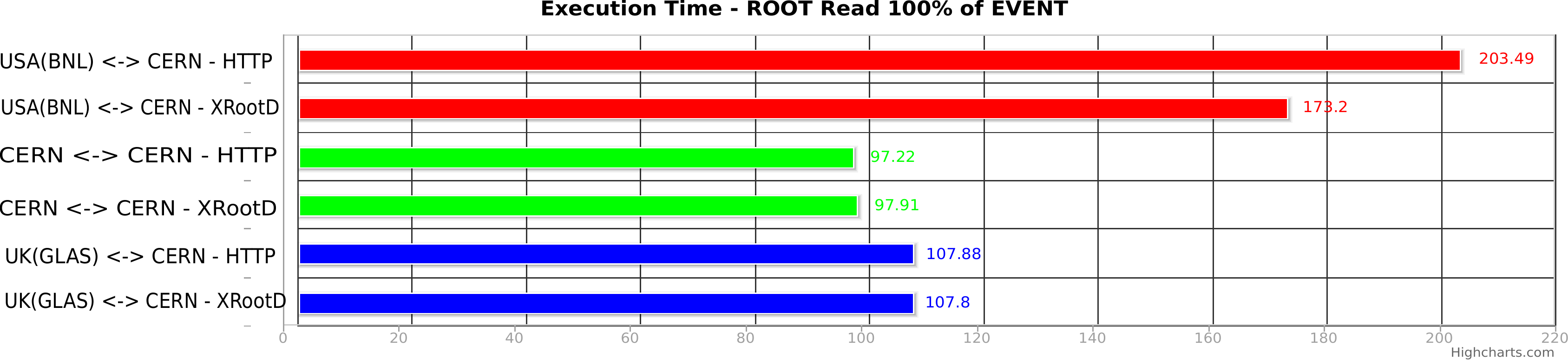}
   \label{fig:perf_100}
   \caption{Execution time of a ROOT data analysis job (less is better).}   
\end{figure}

\par In case of "CERN-CERN" data transfer, the  (Figure-\ref{fig:perf_100}) shows that \textit{libdavix} is respectively 0.7 \%  faster than XRootD in case of local access with high speed link and low latency. This shows that HTTP with \textit{libdavix} can compete with a HPC specific I/O protocol on local area network and offers similar performances in term of data access time and data transfer rate. \\
 
\par In case of "UK(GLAS)-CERN", XRootd and  \textit{libdavix} offers sensibly the same performance. \\
 
\par In case of "USA(BNL)-CERN" data transfer, our tests shows that XRootD is in average 17.5\% faster than \textit{libdavix} on Wide Area network links with high latency. This difference of performance comes mainly from the sliding windows buffering algorithm of XRootD which allows to minimize the number of network round trips executed. Network round trips are naturally extremely costly on high latency networks. \\ 

\par In a classical High Energy Physics grid computing model, a job is always sent close to the data that it will process. Data access are in this case made over a LAN with high speed and low latency. \\
In such model, these results are particularly encouraging for \textit{libdavix} and HTTP I/O.

\section{Conclusion}
Our results shows that for a HPC I/O use case, our solution, \textit{libdavix} can provide similar performance over low latency link to a HPC specific protocol like XRootD. \\
The usage of the HTTP multi-range feature allows to reduce drastically the number of parallel I/O operations and allows HTTP to compete with the aggressive caching strategy of the HPC specific protocols in case remote partial I/O operations on large data sets. \\
The lack of multi-plexing support in HTTP can be compensated by a session recycling system for HEP uses cases and allows to be retro-compatible with the existing network and service infrastructure.  \\
Finally, the association of the Metalink with HTTP gives new possibilities for transparent error recovery in HPC data access and offers an interesting alternative to classical hierarchical data federations. 
\par We have demonstrated in this paper with \textit{libdavix} that an optimized I/O layer based on the HTTP protocol can be considered as a serious and performant alternative to the common HPC specific I/O protocols. \\

\section{Acknowledgments}
We thank Oliver Keeble, Martin Hellmich, Ivan Calvet and Alejandro Alvarez Ayllon for their contributions, commitment and testing dedicated the davix toolkit. Thank you also to Tigran Mkrtchyan for his contributions to TDavixFile. We also thank the ROOT development team for their support and help during the integration of davix to the ROOT analysis framework. Thank you to Olivier Perrin for his support and advises in this work.

\newpage

\bibliographystyle{abbrv}
\bibliography{davix_vldb2}

\end{document}